\documentclass[twocolumn]{aastex631}
\def \eg {e.g.}

\def \ie {i.e.}

\def \lcdm {{\hbox{$\Lambda$CDM}}}
\def \omegam {{\hbox{$\Omega_{\rm m}$}}}
\def \omegal {{\hbox{$\Omega_\Lambda$}}}
\def \hzero {{\hbox{$H_0$}}}
\def \arcmin {\hbox{$^\prime$}}
\def \arcsec {\hbox{$^{\prime\prime}$}}
\def \deg {\hbox{$^\circ$}}

\def \msun {\hbox{${\rm M_\odot}$}}

\def \mfive {\hbox{$M_{500}$}}

\def \rfive {\hbox{$r_{500}$}}

\newcommand{\kmsmpc }{\mbox{km s$^{-1}$ Mpc$^{-1}$}}

\newcommand{\mujyb }{\mbox{$\mu$Jy beam$^{-1}$}}
\newcommand{\mujyarcsecsq }{\mbox{$\mu$Jy arcsec$^{-2}$}}
\newcommand{\muG }{\mbox{$\mu$G}}
\newcommand{\whz }{\mbox{W Hz$^{-1}$}}

\newcommand{\obsid }{ObsID}

\newcommand{\uv }{\textit{uv}}

\newcommand{\wsclean }{\textsc{WSClean}}

\newcommand{\sas }{\textsc{sas}}
\newcommand{\sasE }{Scientific Analysis System}

\newcommand{\halofdca }{\textsc{Halo-FDCA}}
\newcommand{\halofdcaE }{Halo-Flux Density CAlculator}

\newcommand{\xmm }{{\em XMM-Newton}}

\newcommand{\planck }{{\em Planck}}

\newcommand{\rosat }{ROSAT}

\newcommand{\erosita }{eROSITA}

\newcommand{\gmrt }{GMRT}
\newcommand{\gmrtE }{Giant Metrewave Radio Telescope}

\newcommand{\vla }{VLA}
\newcommand{\vlaE }{Very Large Array}

\newcommand{\lofar }{LOFAR}
\newcommand{\lofarE }{LOw Frequency ARray}
\newcommand{\ska }{SKA}
\newcommand{\skaE }{Square Kilometer Array}

\newcommand{\lotss }{LoTSS}
\newcommand{\lotssE }{LOFAR Two-meter Sky Survey}

\newcommand{\nvss }{NVSS}
\newcommand{\nvssE }{NRAO VLA Sky Survey}

\newcommand{\rass }{RASS}
\newcommand{\rassE }{ROSAT All-Sky Survey}

\newcommand{\mcxcE }{Meta-Catalogue of X-ray detected Clusters of galaxies}

\usepackage{graphicx}
\usepackage{amssymb}
\usepackage{multirow,bigdelim}
\usepackage{txfonts}
\shorttitle{The Ant Cluster}
\shortauthors{Botteon et al.}

\begin{document}

\title{Discovery of a radio halo (and relic) in a $\mfive < 2 \times 10^{14}$ \msun\ cluster}

\correspondingauthor{Andrea Botteon}
\email{botteon@strw.leidenuniv.nl}

\author[0000-0002-9325-1567]{A. Botteon}
\affiliation{Leiden Observatory, Leiden University, PO Box 9513, 2300 RA Leiden, The Netherlands}

\author[0000-0003-4046-0637]{R. Cassano}
\affiliation{INAF - IRA, via P.~Gobetti 101, 40129 Bologna, Italy}

\author[0000-0002-0587-1660]{R. J. van Weeren}
\affiliation{Leiden Observatory, Leiden University, PO Box 9513, 2300 RA Leiden, The Netherlands}

\author[0000-0001-5648-9069]{T. W. Shimwell}
\affiliation{ASTRON, the Netherlands Institute for Radio Astronomy, Postbus 2, 7990 AA Dwingeloo, The Netherlands}
\affiliation{Leiden Observatory, Leiden University, PO Box 9513, 2300 RA Leiden, The Netherlands}

\author[0000-0002-5068-4581]{A. Bonafede}
\affiliation{Dipartimento di Fisica e Astronomia, Universit\`{a} di Bologna, via P.~Gobetti 93/2, 40129 Bologna, Italy}
\affiliation{INAF - IRA, via P.~Gobetti 101, 40129 Bologna, Italy}

\author[0000-0002-3369-7735]{M. Br\"{u}ggen}
\affiliation{University of Hamburg, Hamburger Sternwarte, Gojenbergsweg 112, 21029 Hamburg, Germany}

\author[0000-0003-4195-8613]{G. Brunetti}
\affiliation{INAF - IRA, via P.~Gobetti 101, 40129 Bologna, Italy}

\author[0000-0003-4454-132X]{V. Cuciti}
\affiliation{University of Hamburg, Hamburger Sternwarte, Gojenbergsweg 112, 21029 Hamburg, Germany} 

\author[0000-0003-1246-6492]{D. Dallacasa}
\affiliation{Dipartimento di Fisica e Astronomia, Universit\`{a} di Bologna, via P.~Gobetti 93/2, 40129 Bologna, Italy}
\affiliation{INAF - IRA, via P.~Gobetti 101, 40129 Bologna, Italy}

\author[0000-0003-4439-2627]{F. de Gasperin}
\affiliation{Dipartimento di Fisica e Astronomia, Universit\`{a} di Bologna, via P.~Gobetti 93/2, 40129 Bologna, Italy}
\affiliation{University of Hamburg, Hamburger Sternwarte, Gojenbergsweg 112, 21029 Hamburg, Germany}

\author[0000-0002-8648-8507]{G. Di Gennaro}
\affiliation{Leiden Observatory, Leiden University, PO Box 9513, 2300 RA Leiden, The Netherlands}

\author[0000-0002-9112-0184]{F. Gastaldello}
\affiliation{INAF - IASF Milano, via A.~Corti 12, 20133 Milano, Italy}

\author[0000-0002-8286-646X]{D. N. Hoang}
\affiliation{University of Hamburg, Hamburger Sternwarte, Gojenbergsweg 112, 21029 Hamburg, Germany} 

\author[0000-0002-9775-732X]{M. Rossetti}
\affiliation{INAF - IASF Milano, via A.~Corti 12, 20133 Milano, Italy}

\author[0000-0001-8887-2257]{H. J. A. R\"{ottgering}}
\affiliation{Leiden Observatory, Leiden University, PO Box 9513, 2300 RA Leiden, The Netherlands}



\begin{abstract}
\noindent
Radio halos are diffuse synchrotron sources observed in dynamically unrelaxed galaxy clusters. Current observations and models suggest that halos trace turbulent regions in the intra-cluster medium where mildly relativistic particles are re-accelerated during cluster mergers. Due to the higher luminosities and detection rates with increasing cluster mass, radio halos have been mainly observed in massive systems ($\mfive \gtrsim 5 \times10^{14}$ \msun). Here, we report the discovery of a radio halo with a largest linear scale of $\simeq$750 kpc in PSZ2G145.92-12.53 ($z=0.03$) using \lofar\ observations at 120$-$168 MHz. With a mass of $\mfive = (1.9\pm0.2) \times 10^{14}$ \msun\ and a radio power at 150~MHz of $P_{150} = (3.5 \pm 0.7) \times 10^{23}$ \whz, this is the least powerful radio halo in the least massive cluster discovered to date. Additionally, we discover a radio relic with a mildly convex morphology at $\sim$1.7 Mpc from the cluster center. Our results demonstrate that \lofar\ has the potential to detect radio halos even in low-mass clusters, where the expectation to form them is very low ($\sim$5\%) based on turbulent re-acceleration models. Together with the observation of large samples of clusters, this opens the possibility to constrain the low end of the power-mass relation of radio halos.
\end{abstract}


\section{Introduction}

Radio halos are extended synchrotron sources with steep spectra ($\alpha > 1$, with $S_\nu \propto \nu^{-\alpha}$ where
$S_\nu$ is the flux density at frequency $\nu$ and $\alpha$ is the spectral index) that are often observed in massive merging galaxy clusters \citep[\eg][for a review]{vanweeren19rev}. Their emission generally follows the distribution of the thermal gas of the intra-cluster medium (ICM), and occupies the central volume of the cluster. Nowadays, turbulent re-acceleration is thought to be the main mechanism responsible for generating radio halos \citep[\eg][for a review]{brunetti14rev}. In this scenario, relativistic electrons in halos are re-energized due to the interaction with MHD turbulence that has been injected in the ICM during cluster collisions. The competition between energy losses and acceleration of particles (that is connected to the energetics of the merger) results in a gradual steepening of the halo spectrum at higher frequencies \citep[\eg][]{cassano05}. This implies that, especially when observed at GHz-frequencies, they should preferentially be found in massive objects undergoing energetic merging events and be rarer in less massive merging-systems \citep{cassano06}. \\
\indent
The radio halo power is found to increase with the cluster mass and/or X-ray luminosity \citep[\eg][]{feretti12rev, vanweeren19rev}, implying that it is more challenging to observe radio halos in lower mass systems. For this reason, \gmrtE\ (\gmrt) and \vlaE\ (\vla) observations have been focused mainly on massive ($\mfive \gtrsim 5 \times10^{14}$ \msun) or X-ray bright ($L_{\rm X} \gtrsim 5 \times 10^{44}$ erg s$^{-1}$), nearby ($z\lesssim0.4$), clusters \citep{venturi07, giovannini09, cassano13, kale13, kale15, cuciti15, cuciti21a, cuciti21b}. However, one major expectation of the re-acceleration scenario is that the occurrence of radio halos should be higher at low frequencies, where also radio halos generated during less energetic events, \ie\ in low-mass clusters or in minor mergers, should be visible \citep{cassano06, cassano10lofar}. The \lofarE\ (\lofar) is enabling highly sensitive observations at $<200$ MHz which allow us to probe the population of radio halos in the relatively poorly explored regimes of high-$z$ \citep{cassano19, digennaro21fast} and low-mass systems \citep{botteon19lyra, hoang19a2146, hoang21a990, osinga21}. \\
\indent
In this Letter, we report on the discovery of a radio halo (and relic) in PSZ2G145.92-12.53. This cluster belongs to the Clusters in the Zone of Avoidance sample \citep[CIZA;][]{ebeling02} and to the second \planck\ Sunyaev-Zel'dovich (SZ) catalog \citep[PSZ2;][]{planck16xxvii}, is located at $z=0.03$ \citep{ebeling02}, and has a SZ-derived mass of $\mfive = (1.9\pm0.2) \times 10^{14}$ \msun, consistent with that reported by PSZ1 ($\simeq1.8 \times 10^{14}$ \msun; \citealt{planck14xxix}) and in the \mcxcE\ ($\simeq 1.9 \times 10^{14}$ \msun; \citealt{piffaretti11}). Owing to its low-mass and (radio) power, we nickname it the Ant Cluster. Throughout this work, we adopt a \lcdm\ cosmology with $\omegal = 0.7$, $\omegam = 0.3$ and $\hzero = 70$ \kmsmpc. At $z=0.03$, this corresponds to a luminosity distance of $D_{\rm L} = 131.4$ Mpc and to an angular scale of 36.1 kpc arcmin$^{-1}$.

\begin{figure*}[t]
 \centering
 \begin{tabular}{cc}
 \hspace{-1.5cm}
 \multirow{2}{*}{{\includegraphics[width=.65\hsize,trim={0cm 0.0cm 0.0cm 0.0cm},clip]{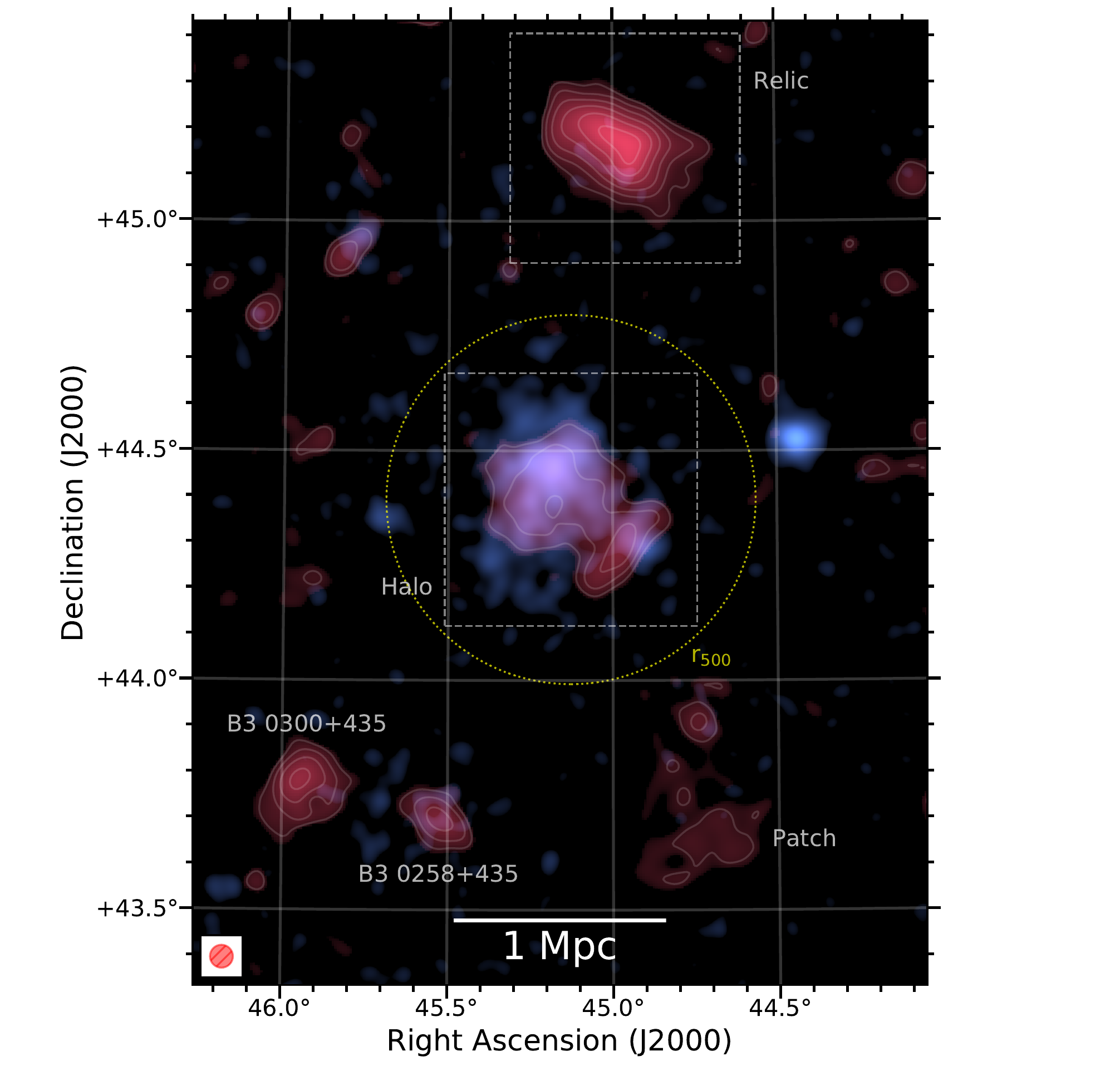}}} & \vspace{-0.35cm} \\
 & \vspace{0.2cm}\hspace{-1.8cm}{\includegraphics[width=.31\hsize]{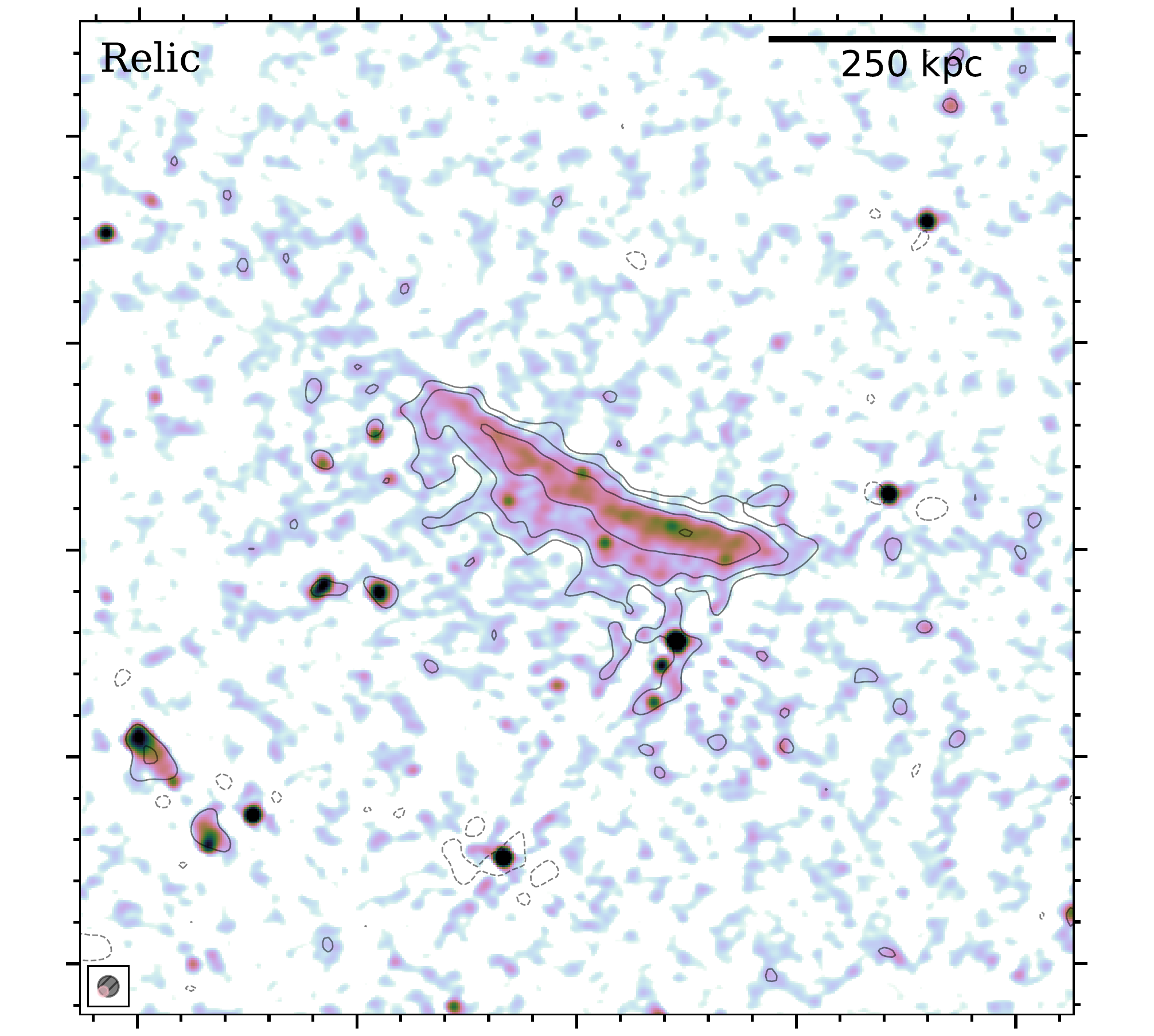}} \\
 & \hspace{-1.8cm}{\includegraphics[width=.31\hsize]{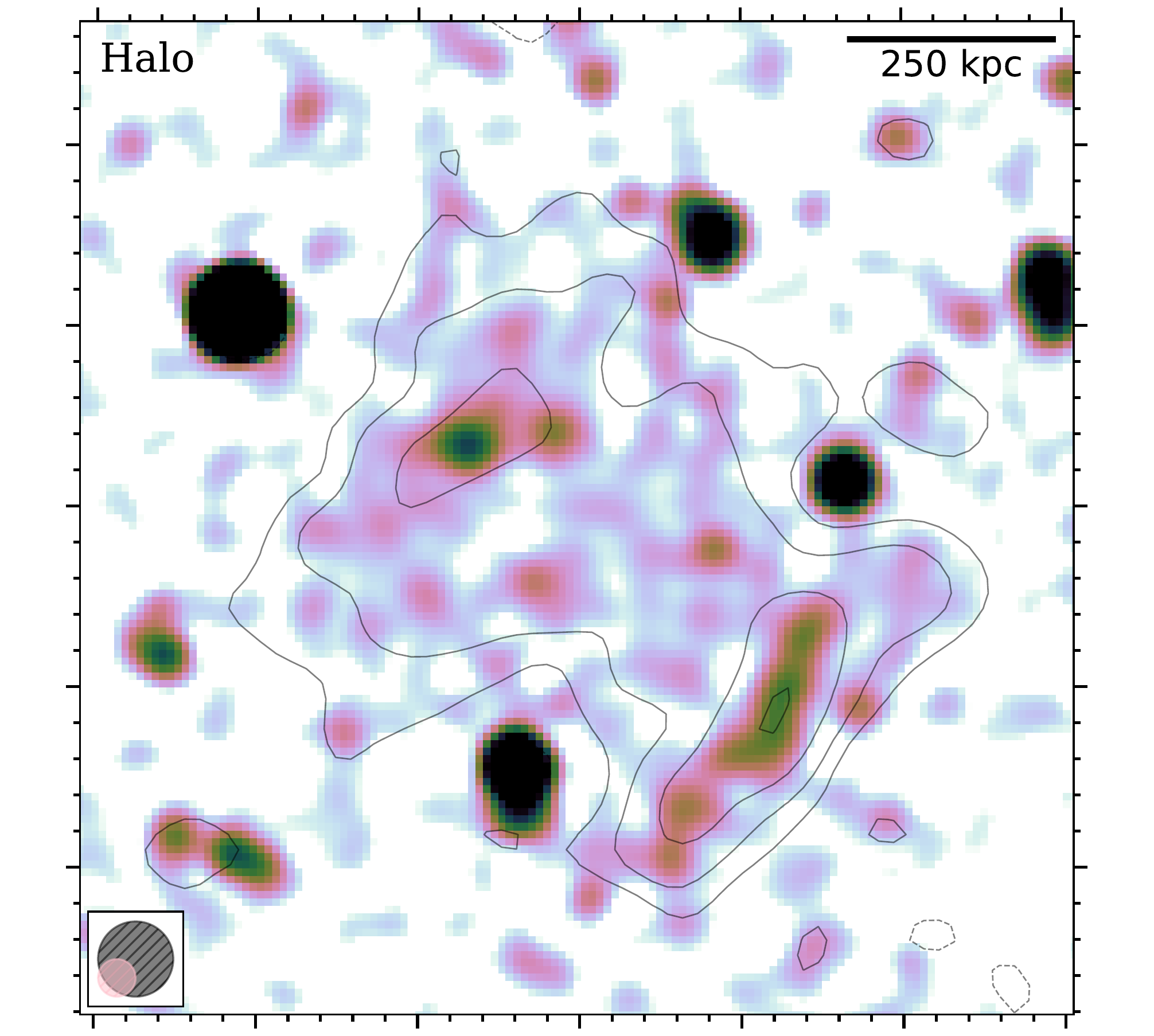}} \\
 \end{tabular}
 \vspace{0.75cm} 
 \caption{\textit{Left panel}: composite radio (\lofar\ 120$-$168 MHz, red) + X-ray (\rosat\ 0.1$-$2.4 keV, blue) image of a $1.6\deg\times2.1\deg$ field centered on the Ant Cluster. The \lofar\ image has a resolution of $180\arcsec$ and was obtained after subtracting discrete sources in the \uv-plane. The dashed boxes mark the regions where we improved the calibration of \lotss\ data. \textit{Right panels}: zoom-in of the (reprocessed) relic and halo regions. Colors indicate the images at 15\arcsec\ and 75\arcsec\ resolution for the relic and halo, respectively. Contours show the discrete source-subtracted emission from images with double beam width. Contours are spaced by a factor of 2 starting from $2.5\sigma$. The $-2.5\sigma$ contour is reported with dashed lines. The beam of the radio images is shown in the bottom left corners. The noise level and resolution are listed in Tab.~\ref{tab:noise}.}
 \label{fig:full}
\end{figure*}

\section{Data reduction}

\subsection{LOFAR}\label{sec:reduction-lofar}

The Ant Cluster is located $\sim$0.95\deg\ from the center of the \lotssE\ \citep[\lotss;][]{shimwell17, shimwell19} pointing P044+44, which was observed for 8~hr on 2018 December 14. Observations were carried out in \texttt{HBA\_DUAL\_INNER} mode adopting a bandwidth of 48~MHz (120$-$168 MHz band), centered at 144~MHz. Preliminary data processing was carried out by the \lofar\ Surveys Key Science Project team adopting the same pipelines used to correct for direction independent and dependent effects across the entire field-of-view (FoV) of \lotss\ observations \citep{vanweeren16calibration, degasperin19, shimwell19, tasse21}. In particular, direction-depended calibration is performed with \texttt{ddf-pipeline}\footnote{\url{https://github.com/mhardcastle/ddf-pipeline}}, which also delivers images of the entire \lofar\ FoV at $6\arcsec$ and $20\arcsec$ resolution. We then obtained images suitable for our study following two approaches. \\
\indent
First, we created a very-low resolution source-subtracted image of a FoV of $1.6\deg\times2.1\deg$ centered on the Ant Cluster. In this case, we used the $20\arcsec$ resolution model of P044+44 obtained from \texttt{ddf-pipeline} to subtract sources in the whole FoV, taking care to exclude any clean components associated with the cluster. The \uv-data were then reimaged at $180\arcsec \times 180\arcsec$ resolution (corresponding to $\simeq 106$ kpc $\times 106$ kpc at the cluster redshift) and corrected for the primary beam response. This image was obtained only for visualization purposes, and was not use further in the analysis. \lotss\ images at $6\arcsec$ and $20\arcsec$ of the same FoV without discrete sources subtracted are shown in Appendix~\ref{app:lotss}. \\
\indent
Second, we refined the calibration towards two regions of interest (halo and relic regions) following the ``extraction and re-calibration'' method described in \citet{vanweeren20arx}. Briefly, this method consists in removing all sources outside a region containing the target from the \uv-data, phase shifting to the center of the region while averaging the data to speed up further processing, correcting for the \lofar\ primary beam response, and performing phase and amplitude self-calibration loops in the extracted region. The improved calibration provided by this approach allows us to study better the cluster diffuse radio emission and to measure its flux density. For this reason, all the flux densities reported in this paper were evaluated using these datasets. Each of the reprocessed regions was reimaged at two different resolutions, with and without discrete sources (modeled and subtracted using the improved datasets). \\
\indent
Our final \lofar\ images are visualized in Fig.~\ref{fig:full} and their noise and resolution are summarized in Tab~\ref{tab:noise}. All images were produced with \wsclean\ v2.10 \citep{offringa14} adopting an inner \uv-cut of $80\lambda$ (corresponding to an angular scale of 43\arcmin) to reduce sensitivity to extended Galactic emission. Uncertainties on the \lofar\ integrated flux densities are dominated by errors in the absolute flux scale, which is conservatively set to 20\%, in line with LoTSS measurements \citep{shimwell19}. The flux scale of the \lofar\ images was aligned adopting the \nvssE\ \citep[\nvss;][]{condon98} derived multiplicative correction factor of 0.876 on \lofar\ data, following \citet{hardcastle21}.

\begin{table}[t]
 \centering
 \caption{Properties of the \lofar\ images shown in Fig.~\ref{fig:full}.}
 \label{tab:noise}
  \begin{tabular}{lccc}
  \hline
  \hline
  Field & Resolution               & Noise    & Discrete sources \\
        & ($\arcsec\times\arcsec$) & (\mujyb) & \\
  \hline
  Full             & $180\times180$ & 1050 & subtracted     \\
  Relic (colors)   & $15\times15$   & 307  & present \\
  Relic (contours) & $30\times30$   & 438  & subtracted     \\
  Halo (colors)    & $75\times75$   & 596  & present \\
  Halo (contours)  & $150\times150$ & 876  & subtracted     \\
  \hline
  \end{tabular}
\end{table}

\subsection{ROSAT and XMM-Newton}

\rosat\ count images in the energy band 0.1$-$2.4 keV in the direction of the Ant Cluster were retrieved together with the corresponding background and exposure maps from the \rassE\ \citep[\rass;][]{voges99} data archive\footnote{\url{https://heasarc.gsfc.nasa.gov/docs/rosat/rass.html}}. These data were combined to produce the count rate image smoothed with a Gaussian kernel with $\sigma = 3$ pixels (1 pixel = 45\arcsec) that is shown in Fig.~\ref{fig:full}. \\
\indent
From the \xmm\ Science Archive\footnote{\url{http://nxsa.esac.esa.int/nxsa-web/}}, we retrieved a pointed observation of the cluster (\obsid: 0744100301, 14~ks). We made use of the \xmm\ \sasE\ (\sas\ v16.1.0) to create clean event files using the tasks \texttt{mos-filter} and \texttt{pn-filter}. 
The MOS1, MOS2, and pn data were then combined to produce a 0.5$-$2.0 keV count rate image that was smoothed with a Guassian kernel with $\sigma = 3$ pixels (1 pixel = 2.5\arcsec). The resulting image, which covers only the central region of the cluster, is shown in Fig.~\ref{fig:overlay}.

\section{Results}

\begin{figure}[t]
 \centering
 \includegraphics[width=\hsize,trim={0cm 0cm 0cm 0cm},clip]{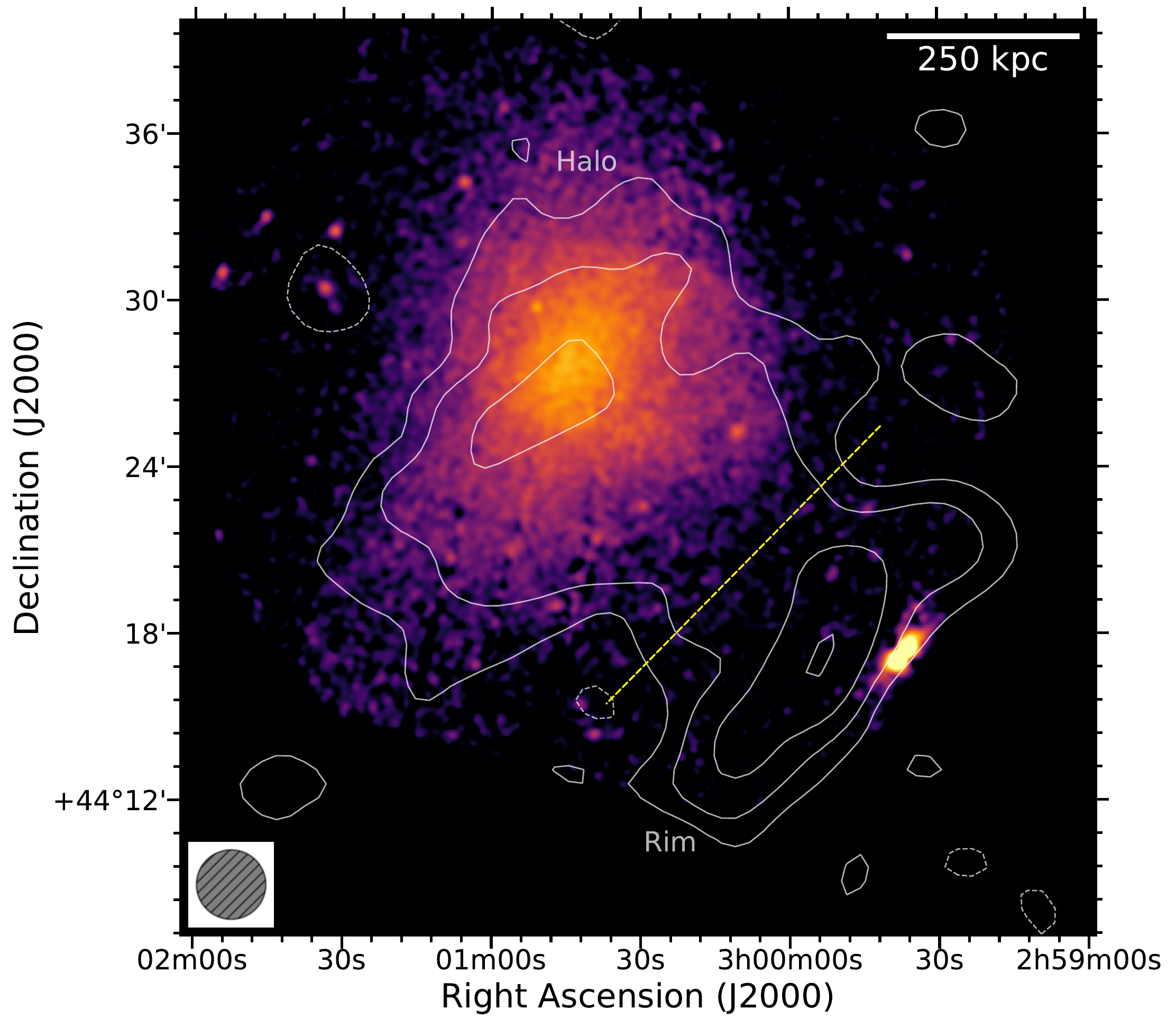}
 \caption{\xmm\ image in the 0.5$-$2.0 keV with overlaid the \lofar\ source-subtracted contours at $150\arcsec$ resolution of Fig.~\ref{fig:full}. The dashed yellow line separates the halo and Rim regions.}
 \label{fig:overlay}
\end{figure}

In the left panel of Fig.~\ref{fig:full}, we show a radio/X-ray (\lotss/\rass) overlay of a large FoV centered on the Ant Cluster. The thermal emission of the ICM (blue) is elongated in the N-S direction, suggesting that the cluster is dynamically unrelaxed. The \lofar\ discrete source-subtracted image (red) highlights the presence of diffuse emission at the center of the cluster that is connected to a brighter and elongated structure to the W, that we term the Rim. At $\sim$1.7 Mpc N from the cluster center, a source extended in the the E-W direction is also detected. Its surface brightness drops rapidly to the N, while it declines more slowly towards the cluster. The zoom-in images of these two regions of interest are depicted in the right panels of Fig.~\ref{fig:full}. \\
\indent
Due to the morphology, location in the cluster, and largest linear size of $\simeq$750 kpc, we classify the central diffuse emission as a radio halo. This classification is supported by the morphological similarity between the X-ray and radio emission, that can be better observed in Fig.~\ref{fig:overlay} where we overlay the \lofar\ contours onto the \xmm\ image of the cluster center. The absence of a compact and peaked X-ray core and the elongated X-ray morphology revealed in the higher resolution \xmm\ data is indicative of a merging system, confirming the impression from the \rass\ image. The hypothesis that the radio emission in the Ant Cluster is tracing a mini-halo (\ie\ another kind of diffuse emission that can be found at the center of galaxy clusters) is rejected due to: its large linear size, the lack of a powerful radio galaxy at its center, and the absence of a cool-core in the system. The Rim, roughly separated from the halo by the dashed line in Fig.~\ref{fig:overlay}, is located in a region of lower gas density, in a direction where the X-ray emission from the ICM has a protuberance. The total flux density of the diffuse emission measured within the $2.5\sigma$ low-resolution level contour is $253\pm51$ mJy, divided into $\simeq$155 mJy for the halo and $\simeq$98 mJy for the Rim. We also measure the flux density of the radio halo by fitting its surface brightness profile in 2D using the \halofdcaE\footnote{\url{https://github.com/JortBox/Halo-FDCA}} \citep[\halofdca;][]{boxelaar21} and adopting an exponential profile \citep[see also][]{murgia09}. Due to the elongation of the halo, we assumed an elliptical model, which has 6 free parameters: the coordinates of the center ($x_0$ and $y_0$), the central brightness ($I_0$), two \textit{e}-folding radii ($r_1$ and $r_2$), and a rotation angle ($\phi$). The Rim region was masked out during the fit. The relevant best-fit physical quantities of the model are $I_0 = 0.62 \pm 0.04$ \mujyarcsecsq, $r_1 = 166\pm12$ kpc and $r_2 = 122\pm9$ kpc, and the model appears to provide an accurate description of the data ($\chi^2/{\rm d.o.f} = 376.07/377$). The flux density at 150 MHz obtained by integrating this model up to three times the \textit{e}-folding radii (as suggested by \citealt{murgia09}) and assuming a spectral index value of $\alpha=1.5\pm0.3$ (which covers the typical range observed for radio halo spectra, \eg\ \citealt{feretti12rev, vanweeren19rev}) is $S_{150} = 167 \pm 35$ mJy, consistent within uncertainties with that measured in the $2.5\sigma$ contour. Hereafter, we adopt this flux density value for the radio halo, which leads to a radio power at 150 MHz of $P_{150} = (3.5 \pm 0.7) \times 10^{23}$ \whz.

We classify the peripheral emission in the cluster N outskirts as a radio relic. Radio relics are diffuse sources observed in merging clusters that trace synchrotron emitting electrons (re)accelerated at merger shocks. The source position in the system, its major axis elongated perpendicularly with respect to the main elongation of the ICM thermal emission (which generally indicates the merger axis), its brightness distribution declining towards the cluster center, and the absence of a clear optical counterpart support this classification. Its flux density (measured within the $2.5\sigma$ low-resolution contour) is $275\pm55$ mJy, corresponding to a power of $(5.7 \pm 1.1) \times 10^{23}$ \whz. We note that the relic can be also observed with low-significance in the \nvss\ at 1.4~GHz; however, the limited sensitivity and poor resolution of \nvss\ does not allow us to provide a meaningful constraint on its spectral index. \\
\indent
Finally, we note the presence of other three prominent sources in the large FoV image of the Ant Cluster (marked Fig.~\ref{fig:full}). While the emissions labeled B3\,0300+435 and B3\,0258+435 \citep{ficarra85} may well be residuals left over by the subtraction of these two bright and extended sources, we exclude that the Patch is an artifact because of the lack of sources in that region in the \lotss\ 6\arcsec\ resolution image of P044+44 (see Appendix~\ref{app:lotss}). This emission arises only at very low-resolution ($\gtrsim$150\arcsec), is detected just above the $2\sigma$ level, and possibly shows an elongation towards the cluster center.

\section{Discussion}

\subsection{Radio halo}

The Ant Cluster is the least massive cluster known to host a radio halo, allowing us to probe non-thermal phenomena in the low-mass end of the cluster mass function. So far there are only a handful of central, extended, diffuse sources detected in clusters with mass $\mfive \lesssim 4 \times 10^{14}$ \msun\ (\eg\ A3562, \citealt{venturi00, venturi03, giacintucci05}; A2061, \citealt{rudnick09}; PSZ1G018.75+23.57, \citealt{bernardi16kat7}; A2146, \citealt{hlavaceklarrondo18, hoang19a2146}; RXCJ1825.3+3026, \citealt{botteon19lyra}; A1775, \citealt{botteon21a1775}) and no radio halos have been claimed below $\mfive < 2 \times 10^{14}$ \msun. \lofar\ is increasing the number of radio halos observed in low-mass systems thanks to the wide area covered by  \lotss-DR2 (Shimwell et al., in prep.) and the Deep Fields observations \citep{osinga21}. Currently, the Ant Cluster is the clearest halo in the lowest mass PSZ2 cluster that we have found after searching \lotss\ images spanning $\sim$35\% of the Northern sky. \\
\indent
In the context of turbulent re-acceleration, radio halos originate during cluster-cluster mergers where a small fraction of gravitational energy is dissipated by large-scale motions into particle acceleration on much smaller scales \citep[\eg][]{brunetti14rev}. The energetic of the merger depends on the mass of the colliding clusters \citep[\eg][]{sarazin02rev}, thus the formation of observable levels of synchrotron emitting electrons is favored in massive systems. This is reflected in the observed steep correlation between radio halo power and cluster mass \citep[\eg][]{cassano13} and in the observed increase of the occurrence of radio halos with the cluster mass \citep[\eg][]{cuciti21b}. Both these observational facts are probed for massive clusters ($\mfive \gtrsim 5 \times 10^{14}$ \msun) and at GHz-frequencies. 

\begin{figure}[t]
 \centering
 \includegraphics[width=\hsize,trim={0cm 0cm 0cm 0cm},clip]{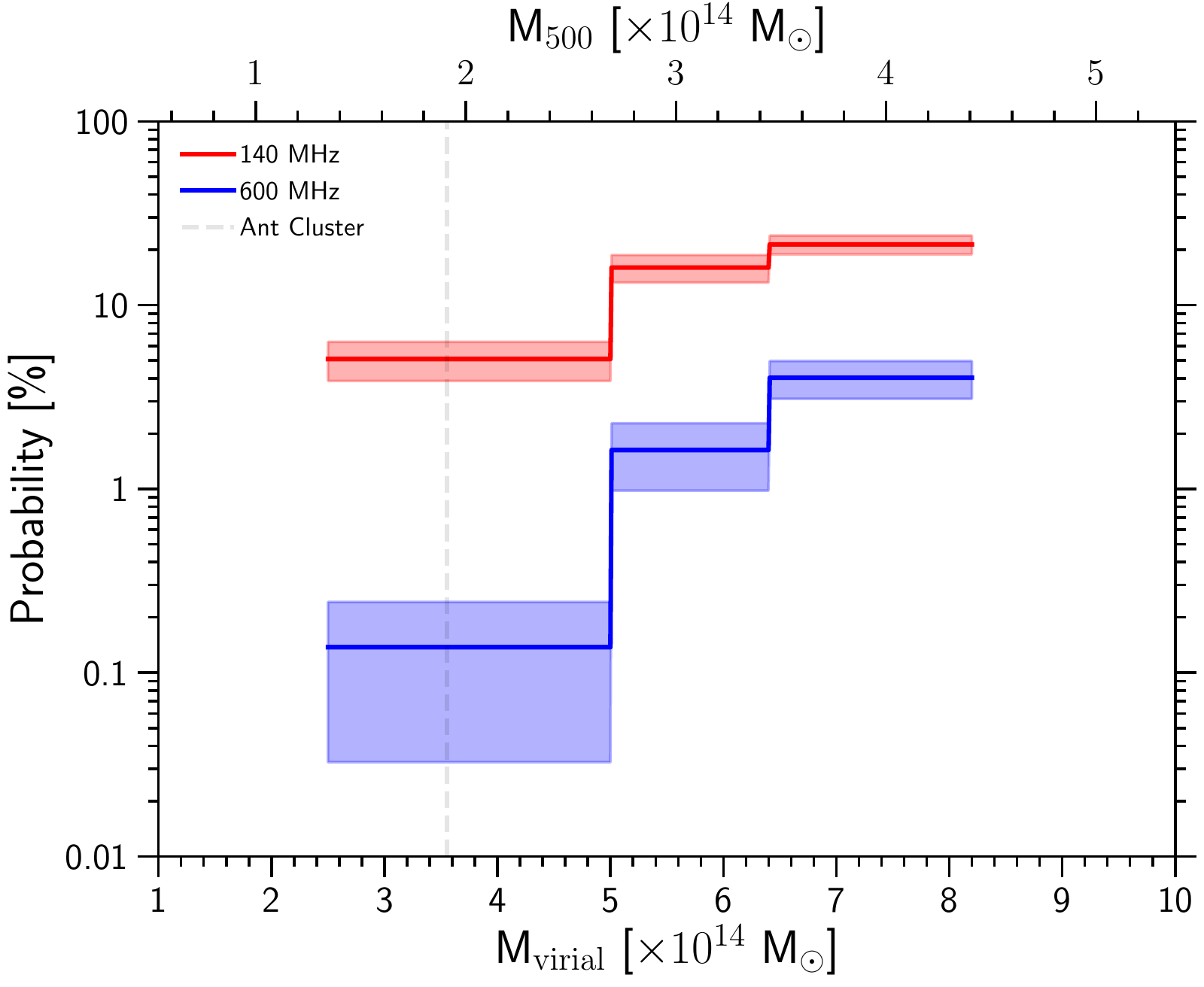}
 \caption{Probability of forming radio halos with $\nu_{\rm s} \gtrsim 140$ MHz and $\nu_{\rm s} \gtrsim 600$ MHz (red and blue lines, respectively) as a function of the cluster mass in the redshift range $z=0-0.1$. A magnetic field of $B=2$ \muG\ is assumed. The shadowed regions represent the $1\sigma$ uncertainty derived through Monte Carlo calculations. The dashed vertical line indicates the mass of the Ant Cluster.}
 \label{fig:occurrence}
\end{figure}

According to re-acceleration models, the occurrence of radio halos should increase at lower frequency especially for less massive systems due to the presence of ultra steep-spectrum radio halos \citep[\eg][]{cassano06, cassano10lofar}. As spectral steepening in radio halos makes it difficult to detect them at frequencies higher than the frequency $\nu_{\rm s}$ at which the steepening becomes severe, one can assume that a radio halo can be successfully detected at a given frequency $\nu_0$ only when $\nu_{\rm s} > \nu_0$. To derive the expected probability to form a radio halo with a given $\nu_{\rm s}$ we use the statistical model developed in \citet[][see also \citealt{cassano06, cassano10lofar}]{cassano05}, which is based on a Monte Carlo approach to describe the merger history of galaxy clusters, to calculate the generation of turbulence, the particle acceleration, and the synchrotron spectrum during the cluster lifetime. In Fig.~\ref{fig:occurrence} we show the probability to form a radio halo as a function of the cluster mass, in the redshift range $z=0-0.1$, with a steepening frequency $\nu_{\rm s} \gtrsim 140$ MHz and $\nu_{\rm s} \gtrsim 600$ MHz. We find that, for a cluster with a mass similar to that of the Ant Cluster, the probability to form a radio halo with steepening frequency $\nu_{\rm s} \gtrsim 140$ MHz is $\sim$5\% (red line), while it decreases down to a few per thousand for $\nu_{\rm s} \gtrsim 600$ MHz (blue line), with a clear dependence on cluster mass. The large difference expected between the two fractions implies that a large number of these halos should have very steep radio spectra. The calculations assume a mean rms magnetic field average over the radio halo volume of $B=2$ \muG. Decreasing the magnetic field to $B=1$ \muG\ has only a small impact in the formation probability at 140~MHz (well within the reported uncertainties), while a more important decrement is expected at 600~MHz. These $B$ values have been chosen to be in line with the magnetic field strengths at $\sim$\muG\ levels observed in galaxy clusters \citep[\eg][for a review]{govoni04rev}.

\begin{figure}[t]
 \centering
 \includegraphics[width=\hsize,trim={0cm 0cm 0cm 0cm},clip]{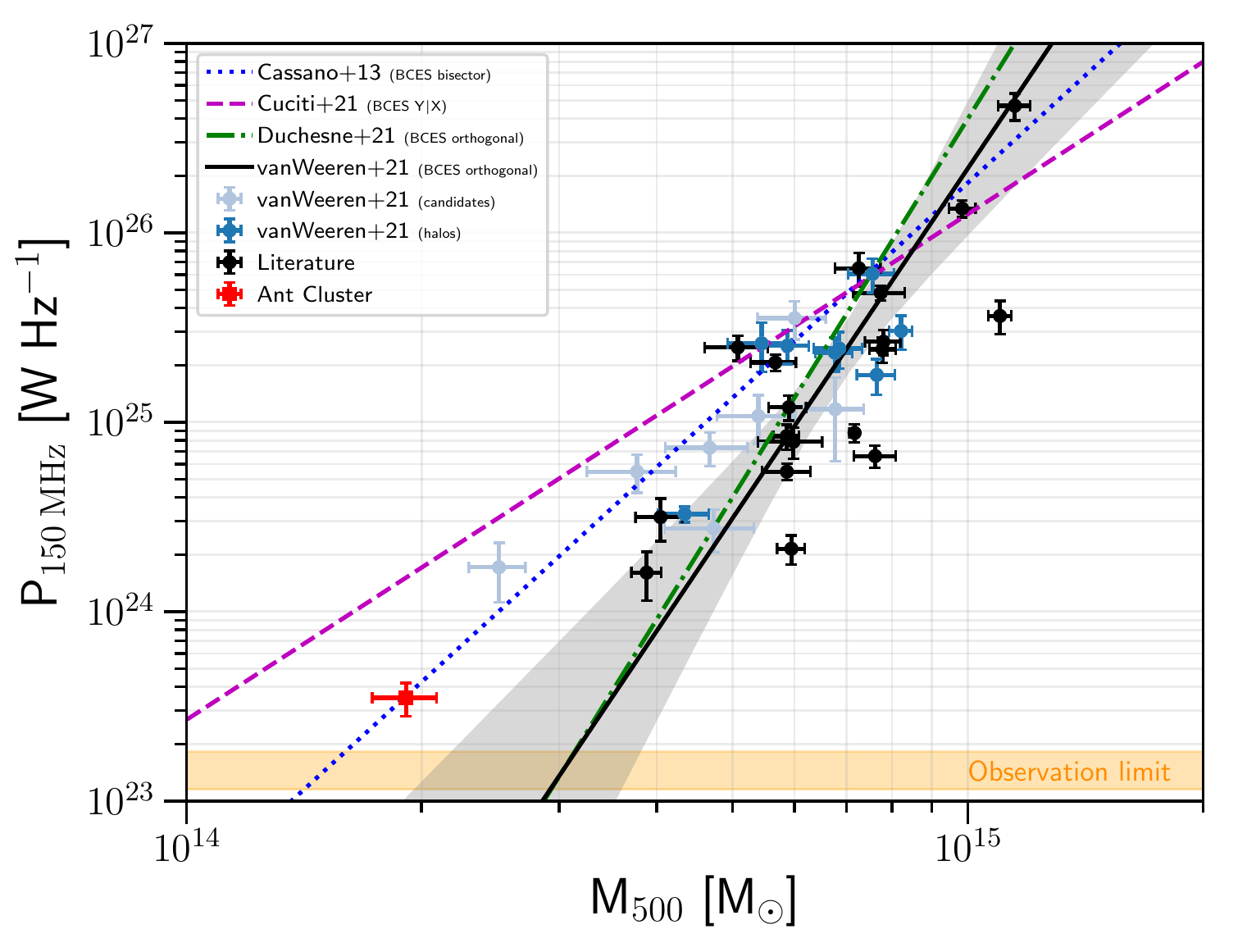}
 \caption{The position of the Ant Cluster in the $P_{150}-\mfive$ relation of \citet{vanweeren20arx}. The reference scaling relations reported by \citet{cassano13, cuciti21b, duchesne21eor}, rescaled to 150~MHz adopting $\alpha=1.3$, are also shown. The orange horizontal region marks the detection limit for our \lofar\ $180\arcsec$ image.}
 \label{fig:relation}
\end{figure}

In Fig.~\ref{fig:relation} we plot the position of the Ant Cluster in the $P_{150}-\mfive$ relation from \citet{vanweeren20arx}. The Ant Cluster is the least powerful radio halo observed to date, and lays in a region were we need to extrapolate the known relations. Its position on the diagram shows that with \lofar\ we are entering in an unexplored regime of cluster mass. Only with the analysis of large statistical samples of clusters, taking into account possible selection effects (\eg\ completeness in mass) and observational biases (\eg\ sensitivity of radio observations), it will be possible to firmly constrain the slope of the $P_{150}-\mfive$ relation.  \\
\indent
Future X-ray and radio surveys performed with \erosita\ \citep{predehl21} and the \skaE\ \citep[\ska;][]{dewdney09} are expected to discover many new low-mass clusters missed by previous instruments, allowing us to definitely enter into the study of non-thermal phenomena in this poorly explored regime of cluster mass.

\subsection{Relic, Rim, and Patch}

To the best of our knowledge, the Ant Cluster is the third cluster with an SZ-derived mass $\mfive < 2\times10^{14}$ \msun\ that hosts a radio relic (the others are A168, \citealt{dwarakanath18}; and A1904, \citealt{vanweeren20arx}). The relic is located at a considerable distance from the cluster center ($\sim$ 1.7 Mpc $\simeq$ 2\rfive) and shows a mildly convex morphology (Fig.~\ref{fig:full}, top right panel). Relics are generally observed at $0.5-2$ Mpc from the cluster center \citep[\eg][]{degasperin14, nuza17} and have a concave morphology. The mildly convex curvature of the relic in the Ant Cluster is quite atypical, and may arise from the properties of the medium encountered by the shock front during its long propagation into the cluster outskirts. Many aspects of the formation mechanisms of radio relics are still uncertain, and the discovery of more relics with similar properties will help to refine the models \citep[\eg][]{bruggen20}. \\
\indent
The classification of the Rim and Patch is still not clear. The Rim has an elongated morphology and higher surface brightness compared to the radio halo, possibly suggesting a radio relic nature. We note that this emission is projected within \rfive, where relics should be more unlikely to observed \citep{vazza12why}. The Patch has a very low-surface brightness and potentially shows a trail of emission towards the cluster. However, the low-significance of the detection makes it hard to determine the exact morphology and whether or not it is associated with some optical counterpart. Still, its peripheral location and distance to the cluster center similar to that of the relic may suggest that also the Patch is tracing a radio relic. Obviously, this hypothesis is rather speculative at the moment and deeper, multi-frequency observations are required to understand its nature.

\section{Conclusions}

We have reported the discovery of a radio halo in PSZ2G145.92-12.53 (Ant Cluster) at $z=0.03$ using \lofar\ observations at $120-168$ MHz carried out in the context of \lotss. The halo occupies the central region of the cluster and its morphology follows that of the ICM thermal emission. Its radio power at 150~MHz of $P_{150} = (3.5 \pm 0.7) \times 10^{23}$ \whz\ and mass of $\mfive = (1.9\pm0.2) \times 10^{14}$ \msun\ make it the least powerful and least massive system hosting a radio halo known to date. Our observations show the potential of \lofar\ to detect radio halos even in low-mass systems, where the probability to form them in the context of turbulent re-acceleration models is expected to be very low ($\sim$5\%). Future observations of statistical samples of clusters will allow us to test the model predictions and constrain the low-power and low-mass end of the $P_{150}-\mfive$ relation.

\begin{acknowledgments}
We thank the anonymous referee for constructive comments that helped improve the manuscript. 
ABot and RJvW acknowledge support from the VIDI research programme with project number 639.042.729, which is financed by the Netherlands Organisation for Scientific Research (NWO). RC, GB, FG, and MR acknowledge support from INAF mainstream project `Galaxy Clusters Science with LOFAR' 1.05.01.86.05. ABon and DNH acknowledge support from ERC-Stg DRANOEL 714245. VC acknowledges support from the Alexander von Humboldt Foundation.
LOFAR \citep{vanhaarlem13} is the LOw Frequency ARray designed and constructed by ASTRON. It has observing, data processing, and data storage facilities in several countries, which are owned by various parties (each with their own funding sources), and are collectively operated by the ILT foundation under a joint scientific policy. The ILT resources have benefitted from the following recent major funding sources: CNRS-INSU, Observatoire de Paris and Universit\'{e} d'Orl\'{e}ans, France; BMBF, MIWF-NRW, MPG, Germany; Science Foundation Ireland (SFI), Department of Business, Enterprise and Innovation (DBEI), Ireland; NWO, The Netherlands; The Science and Technology Facilities Council, UK; Ministry of Science and Higher Education, Poland; Istituto Nazionale di Astrofisica (INAF), Italy. 
This research made use of the University of Hertfordshire high-performance computing facility and the LOFAR-UK computing facility located at the University of Hertfordshire and supported by STFC [ST/P000096/1]. 
This research made use of APLpy, an open-source plotting package for Python \citep{robitaille12}.
\end{acknowledgments}

\facilities{\lofar, \xmm, \rosat}

\appendix

\section{Images from \lotss}\label{app:lotss}

\begin{figure*}[t]
 \centering
 \includegraphics[width=.48\hsize,trim={0cm 0.0cm 0.0cm 0.0cm},clip]{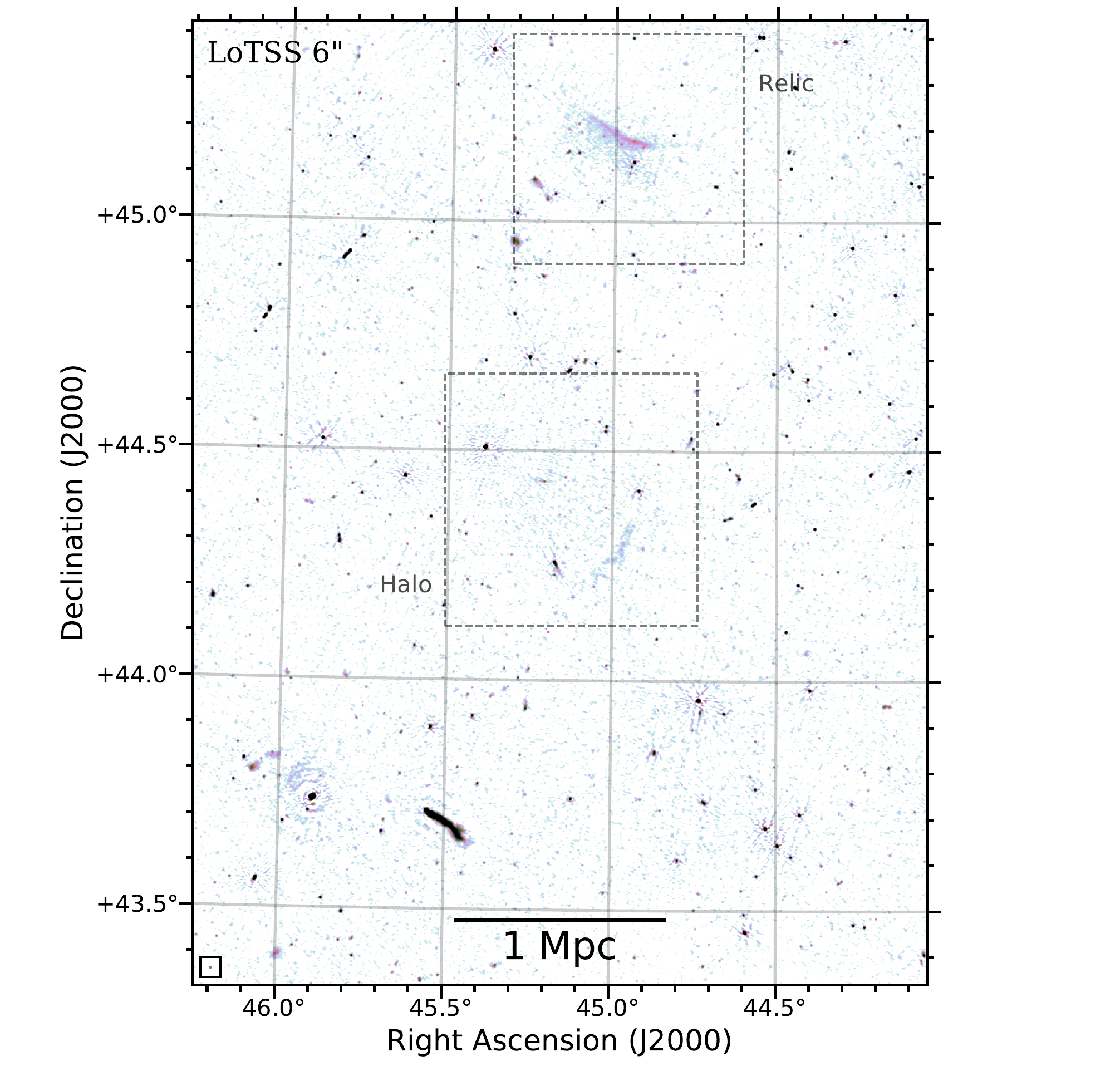}
 \includegraphics[width=.48\hsize,trim={0cm 0.0cm 0.0cm 0.0cm},clip]{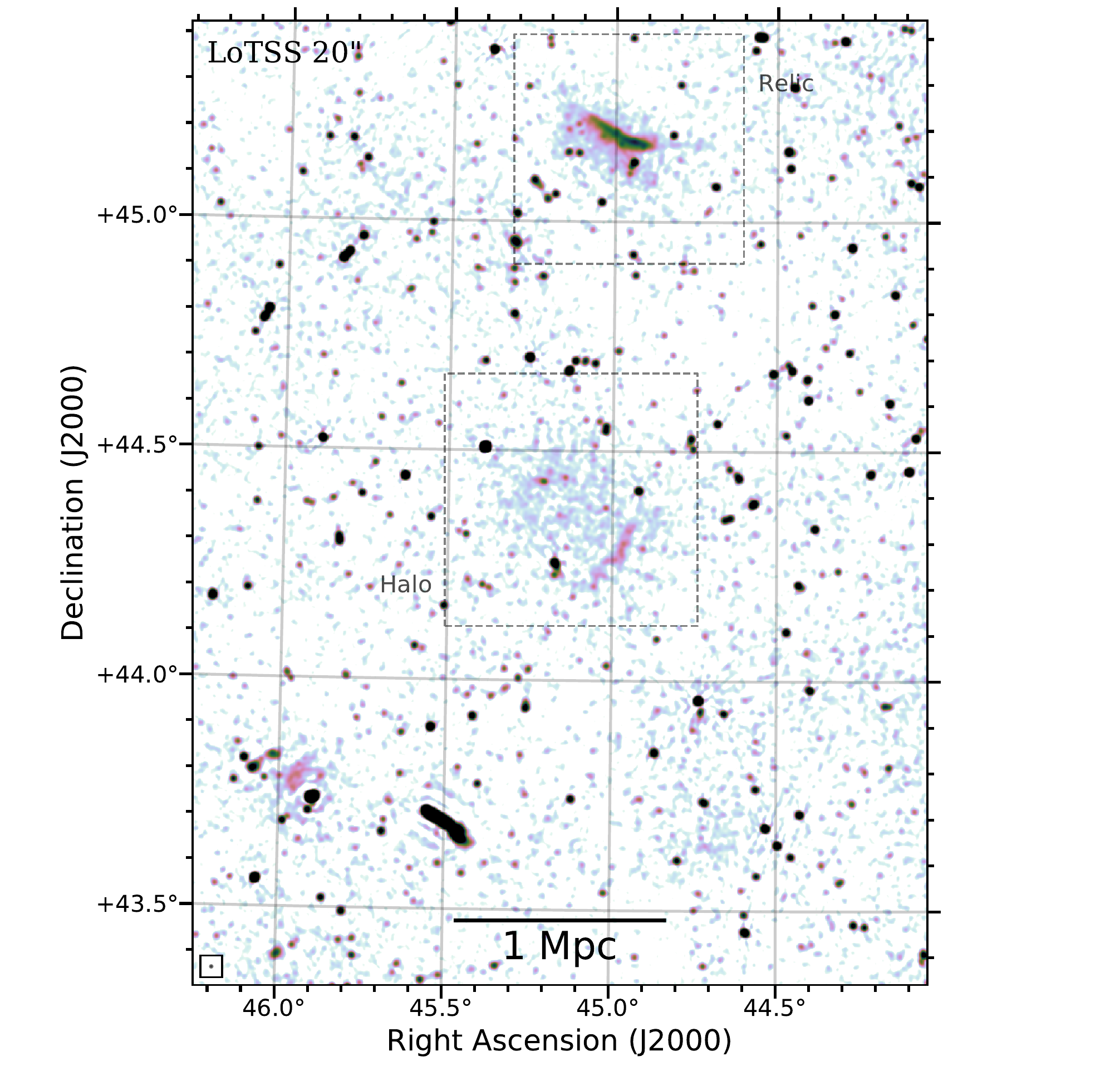}
 \caption{\lotss\ images at $6\arcsec$ (\textit{left}) and $20\arcsec$ (\textit{right}) resolution and with a noise of 93 and 160 \mujyb, respectively. The FoV and dashed boxes are the same of the left panel of Fig.~\ref{fig:full}.}
 \label{fig:lotss}
\end{figure*}

In Fig.~\ref{fig:lotss} we show the corresponding \lotss\ images at $6\arcsec$ and $20\arcsec$ resolution of the FoV covered by the left panel of Fig.~\ref{fig:full} as produced by the \lofar\ Surveys Key Science Project team by running \texttt{ddf-pipeline} on the pointing P044+44. These images were obtained employing the third generation calibration and imaging algorithm described in \citet{tasse21}. Thanks to the ``extraction and re-calibration'' method described in Section~\ref{sec:reduction-lofar}, we are able to: improve the calibration towards specific targets, perform fast and flexible re-imaging of the regions of interest, model and subtract discrete sources more carefully, and deeply deconvolve faint diffuse emission \citep[see][for more details]{vanweeren20arx}.

\bibliographystyle{aasjournal}
\bibliography{library.bib}



\end{document}